# Mixing Grains with Different Elongation in a Rotating Drums

*Claudia Piacenza[1,2], Marco Marconati[1], Colin Hare[1], Andrea C. Santomaso[2], Marco Ramaioli[1,*]*

[1]Department of Chemical and Process Engineering, Faculty of Engineering and Physical Sciences, University of Surrey, Guildford GU2 7XH, UK
[2] APT Lab, Department of Industrial Engineering, University of Padova, Padova, Italy

[*] *Email: m.ramaioli@surrey.ac.uk*

## Abstract

Mixing grains with different properties is a remarkably challenging process, relevant to many industrial applications. Rotating drums have been used extensively as model systems to study granular media flow and mixing and segregation. Numerous studies considered the mixing of grains with different sizes, but only few studies considered shape and elongation, which have been already identified as important characteristics affecting packing and inducing segregation. In this contribution, the mixing of binary mixtures of grains having the same volume, but different elongations was studied experimentally. The mixing dynamics of a layered granular medium was characterized in a rotating drum, highlighting the impact of the grain shape, of the drum angular speed and of the drum filling ratio. A mixed or segregated state is reached very rapidly, but experiments are continued to verify that the state reached is a steady state. The experiments demonstrated that below a critical difference of elongation, the grains can be mixed effectively. Conversely, when grain elongations are very different, a central core is formed, rich in the more elongated grains. In this case only limited mixing can be achieved. These results can guide the formulation of mixtures of grains with different elongations.

## 1. Introduction

Mixing is a process operation aimed at obtaining a homogeneous system starting from two or more components. On the contrary, when there is a lack of homogeneity within the mixture segregation occurs. Segregation phenomena represent a relevant problem in the case of mixing solid particles having different physical properties such as particle size, density, shape or surface roughness. Differently from liquids and gases, solid particles are not thermal activated system [1] and so in order to obtain a certain degree of mixing they need to be mechanically activated. Typically three mixing mechanisms are largely recognized:
- dispersion, analogous to diffusion in fluids, it involves single particles that randomly move due to collision between each other;
- convection, it involves a portion of the bulk material, with a given direction and velocity;
- shear mixing: it is peculiar to granular solids and it is a combination of the two previous mechanisms.

Powder mixers are relatively simple machines of low capital cost. They are classified into two types: tumbling mixers and agitated mixers. Tumbling mixers are the simplest. A totally enclosed vessel is rotated about an axis causing the particles within the mixer to tumble over each other on the mixture surface. Tumbling vessels provide low-shear environments. Convective mixers instead are characterized by the action of impellers or paddles to move the powders around and generate a well-mixed product. Convective mixers provide more shear into the mixture and can be used with cohesive materials.

Rotating drums belong to the tumbling mixers. They are extensively used in the chemical and process industries. With respect to mixing processes, rotating drums are often used to study and understand mixing and segregation phenomena; as a result, many experimental and numerical studies were developed on granular behavior in rotating drums [1-3]. Different granular flow regimes can occur in the drum, each one with its specific flow behavior. In particular, different flow regimes can be assessed by using the dimensionless Froude number, $Fr = \omega^2 R/g$ [4], where $\omega$ [s$^{-1}$] is the drum rotation speed, $R$ [m] the drum radius and $g$ [m/s$^2$] the gravitational acceleration. With $10^{-4}<Fr<10^{-2}$ and filling level $\varphi>0.1$ rolling regime takes place. This motion is characterized by a uniform, steady flow of a particle layer on the surface (cascading or active layer) while the larger part of the bed (plug flow region) is moved upwards by solid rotation with the rotational speed of the wall. A uniform and good mixing is achievable with this type of motion, but when particle size distribution is very broad segregation can appear. The rolling bed is preferred, since is the simplest regime providing favorable conditions for heat transfer and high quality of the product even when mass flow rates are large.

When particles to be mixed have different properties such as different size, different density or different shape segregation typically occurs and particles tend to separate. [5, 6]. Difference in size is by far the most important cause of particle segregation. Particle size distribution influences also strongly bulk properties such as packing fraction and bulk modulus [7].

Different size segregation mechanisms can be identified:
- trajectory mechanism: according to the Stokes' law the velocity of fine particles is lower than that of big particles. This can cause segregation when small particles move through air;
- percolation mechanism: if a bulk material is disturbed, a rearrangement in the particle packing occurs and for fine particles it is easier to fall down in the inter-particle voids;
- inertial mechanism: in pouring and filling operations bigger particles tend to flow selectively to the bottom of the pile since they have a higher momentum;
- elutriation mechanism: in presence of air the very fine particles can remain in suspension while larger ones tends to settled. This can occur in charging storage vessels or hoppers.

A homogeneous binary granular mixture, with differences in size and density, in a two-dimensional rotating drum typically gives radial segregation. Considering a rotating drum operating in the rolling regime filled with particles of different sizes, bigger particles mainly tend to the edge of the system, while smaller particles accumulate to the centre of the drum. The phenomenon is fast and the rotational speed of the drum seems not to affect the phenomenon; the size ratio of the particles is, instead, the parameter that mainly affects size segregation phenomena. Most of the experimental investigations about segregation in the rotating drum involved spherical or nearly-spherical particles; the experimental studies of the flow of non-spherical particles are, in fact, rather limited [6,8]. The shape of the particles, however, strongly influences the motion in the system. In particular, the static and dynamic angles of repose of non-spherical particles are found to be very different from the ones measured for spherical particles suggesting that differences in particle mobility can affect the segregation process [9].

## 2. Materials and methods

### 2.1 Particles

Both spherical and non-spherical particles were used for the experiments. Their size (diameter $d$ for spheres, major and minor axis for ellipsoids $a$ and $b$ respectively), aspect ratio $AR$, projected area $A_p$, volume $V$, and true density $\rho$ are reported in Table 1.

**Table 1.** Physical properties of the particles

| Type | $d$ [mm] | $a$ [mm] | $b$ [mm] | $AR$ [-] | $A_p$ [mm$^2$] | $V$ [mm$^3$] | $\rho$ [g/cm$^3$] |
|---|---|---|---|---|---|---|---|
| Spheres S | 8 | - | - | 1 | 50.3 | 268 | 1.09 |
| Ellipsoids E1 | - | 10.48 | 6.98 | 1.5 | 57.5 | 267 | 0.96 |
| Ellipsoids E2 | - | 18.44 | 5.26 | 3.5 | 76.2 | 267 | 0.98 |

Spherical particles with size 8 mm were used for the experiments. The ellipsoids were 3D printed with *AR* equal to 1.5 and 3.5. They were printed with the Ultimaker 3D printer and made of polylactic acid (PLA). The software used to realize the 3D design of the particles used in this work is FreeCAD, which is an open-source parametric 3D modeller allowing to obtain STL files.

The true density of all the particles used for the experiments was measured by gas pycnometry (Multivolume Pycnometer 1305, Micromeritics). The measured density of the spheres was very close to the typical density of the nylon 66 (1.14 g/cm$^3$). The PLA ellipsoids had a density that was lower than the typical density of the PLA (1.24 g/cm$^3$). This is because the particles were printed with a set infill density of the 20%, so they where partially empty inside.

## 2.2 Tumbling mixer

The quasi-2D drum was made of two transparent polycarbonate disks with a diameter of 30 cm and a thickness of 6 mm; between them a wooden spacer was placed, with the same external diameter, an internal diameter of 25 cm and a thickness of about 9.1 mm. The spacer allowed to create an internal volume of finite thickness separating the two polycarbonate disks.

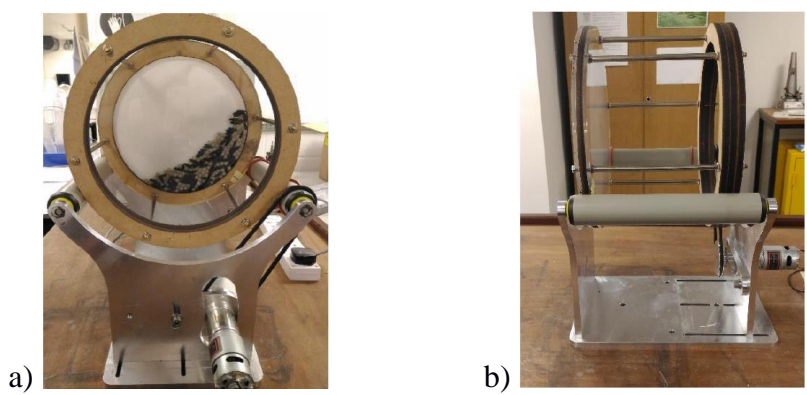

**Figure 1.** a) Front view and b) side view of the rotating drum (the drum is on the left side of the rigid structure).

The internal lateral wall of the volume was covered with sandpaper in order to provide friction and avoid undesired slip of the particles. To keep the drum in vertical position and allow its rotation, three additional wooden spacers were used, at a distance of about 16.3 cm. This rigid structure could rotate at variable rotational speed thanks to a motor connected to a voltage supplier. Two rollers mounted on a rigid support made the rotation of the drum possible; a pulley transmitted the motion to the driving roller, while the second roller was passive. The rotational speed could be changed by varying the motor voltage. The structure could be removed from the rollers and by unscrewing the bolts, the drum could be easily opened so that to load/unload the particles.

## 2.3 Image acquisition

Frontal pictures of the drum were taken by a color camera and analyzed by the image processing software ImageJ [10]. To avoid the reflection of the camera and other objects on the front transparent disk, a squared LED panel was used which allowed to have a dominant diffused light illuminating the drum. The LED had a power of 8 W and a color temperature of 4000 K; it was placed behind the drum and opposite to the camera. A circular sheet of paper, which was attached to the drum between it and the LED, diffused the light properly. The camera was a Basler ace camera − model acA2000-165uc, with a maximum resolution of 2040 × 1086 pixels. The focus and the diaphragm of the camera could be adjusted manually.

## 2.4 Experimental plan

The attention was focused on mixtures of particles with different shapes, since segregation phenomena due to size had been widely investigated in the past. Three mixtures were considered, with two possible initial top-bottom segregated configurations. Filling level and rotational speed where changed also as detailed in Table 2.

**Table 2.** Experimental tests performed with differently shaped particles.

| AR 1 [-] | AR 2 [-] | ω [rpm] | φ [-] | Init. Config. |
|---|---|---|---|---|
| 1 | 1.5 | 4.22 | 25% | E1/S |
| 1 | 1.5 | 4.22 | 25% | S/E1 |
| 1 | 1.5 | 4.22 | 47% | E1/S |
| 1 | 1.5 | 4.22 | 47% | S/E1 |
| 1 | 1.5 | 11.51 | 25% | E1/S |
| 1 | 1.5 | 11.51 | 47% | E1/S |
| 1 | 3.5 | 4.22 | 25% | E2/S |
| 1 | 3.5 | 4.22 | 25% | S/E2 |
| 1 | 3.5 | 4.22 | 47% | E2/S |
| 1 | 3.5 | 11.51 | 25% | E2/S |
| 1 | 3.5 | 11.51 | 25% | S/E2 |
| 1 | 3.5 | 11.51 | 47% | E2/S |
| 1.5 | 3.5 | 4.22 | 25% | E2/E1 |
| 1.5 | 3.5 | 4.22 | 25% | E1/E2 |
| 1.5 | 3.5 | 4.22 | 47% | E2/E1 |
| 1.5 | 3.5 | 4.22 | 47% | E1/E2 |

*AR*1 and *AR*2 indicate the aspect ratios of the particles in the binary mixture. It is reminded that the spheres (*AR*=1) and the two types of ellipsoids (*AR*=1.5 and *AR*=3.5) had the same volume. The E1/S initial configuration indicates a segregated configuration with short ellipsoids above spheres (this was assumed as basis configuration); the S/E1 configuration is inverted with spheres above short ellipsoids. A similar convention is applied to the experiments considering mixtures of ellipsoids. All the experiments were performed with binary mixtures with a composition of 50% and 50% on a number (and hence also volume) basis. The filling level $\varphi$ of the 25% was obtained with a total number of particles equal to 200; with 320 total particles the filling level of the drum was equal to 47%.

## 2.5 Image analysis procedure

Image analysis techniques were used to identify the particles on the basis of their different shape. The procedure was automated using customized ImageJ macros. Contrast and brightness of the images were first adjusted. Color images were then set to gray scale and segmented by applying a threshold. The operation was carried out twice to separate light and dark particles with different thresholds. Watershed segmentation was then used to separate touching objects. This last operation was more difficult for the particles at the border of the drum; in some case, in fact, the ellipsoids were erroneously identified as spheres. Also, the presence of the border of the drum complicated the watershed operation.

The centre of mass of each particle ($x_{m,i}$ and $y_{m,i}$) was estimated and the radial distance ($d_i$) from the centre of the drum ($x_c$, $y_c$) calculated as:

$$d_i=[(x_c-x_{m,i})^2+(y_c-y_{m,i})^2]^{0.5}$$

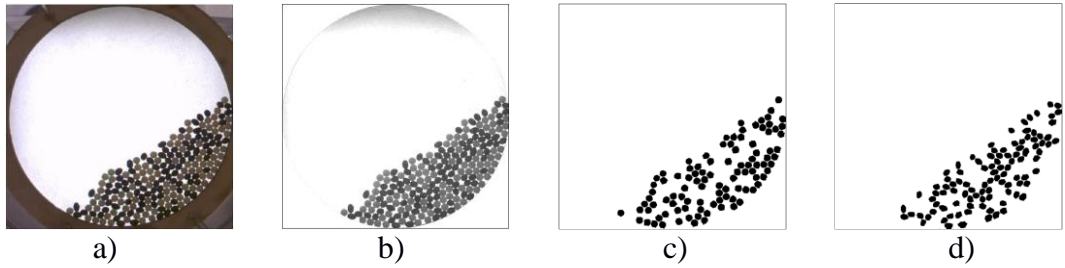

**Figure 2.** Example of image processing steps. a) Original colour image; b) grayscale conversion and circle cut from the original picture; segmentation applied to c) light particles and d) dark particles.

The mean value of distance ($d_m$) for the spheres and for the ellipsoids was then calculated as:
$$d_m = \Sigma\, d_i/N$$
Only for very elongated ellipsoid ($AR=3.5$) the watershed operation failed resulting in over-segmentation errors. Single particles could not be identified unequivocally in this case; only one center of mass for all the elongated ellipsoid was estimated and the average distance ($d_m$) of the particles from the drum center calculated.

## 3. Results and discussion

Figure 3 shows two typical experiments, carried out at the lowest rotational speed and filling ratio, with ellipsoids ($AR=1.5$ at left and $AR=3.5$ at right) on top and spheres on bottom in the initial and final configuration. It is evident even by visual inspection that the final state presented some level of radial segregation and that a different final segregation intensity was reached in the two systems.

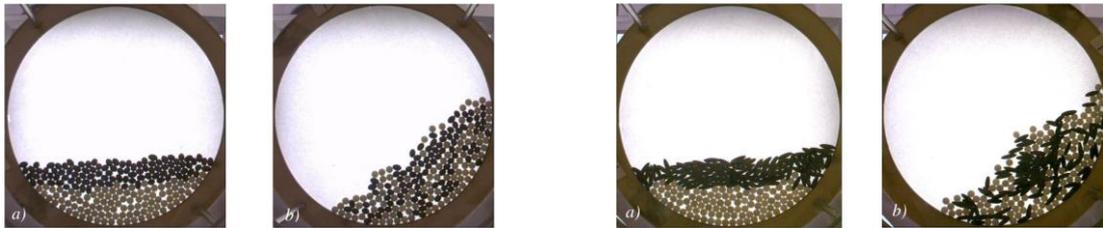

**Figure 3.** Drum at a) t=0 and b) t=30 min. On the left the initial configuration is E1/S (using ellipsoids with $AR=1.5$) and on the right the initial configuration is E2/S ($AR=3.5$). In both experiments $\omega=4.22$ rpm and $\varphi=25\%$.

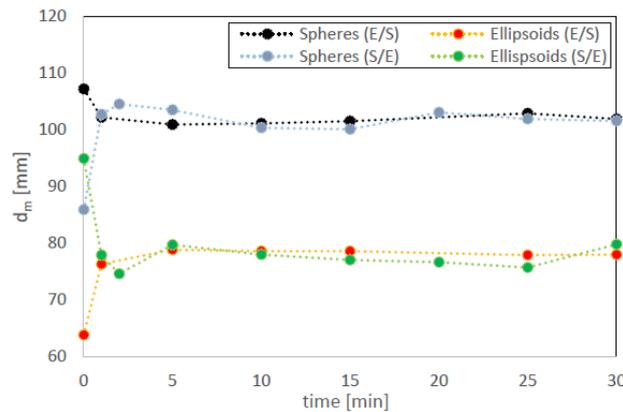

**Figure 4.** Evolution of the mean radial distance vs. time for the mixture of spheres and ellipsoids with $AR=3.5$. The initial configuration E2/S is compared to the initial configuration S/E2.

A more quantitative analysis was obtained by plotting the trend of the mean radial distance ($d_m$) of the two species at different time instants. As a typical example of the system evolution, the case of E2/S is shown in Figure 4. The $d_m$ reached a steady value after a short time. Dedicated experiments on a shorter time scale (2 minutes, not shown here) revealed that steady state value for $d_m$ was reached after 30s for mixtures with ellipsoids E1 ($AR$=1.5) and after 60s for mixtures with E2 ($AR$=3.5).

At steady state, the radial distance was in both cases smaller for the ellipsoids, meaning that they mainly remained concentrated on top or at the center of the granular bed while the spheres mainly concentrated at the periphery of the drum (close to the drum wall). This tendency was much stronger for ellipsoids E2. So a slower segregation dynamics compensated by a larger segregation intensity was observed with the more elongated particles.

As Figure 4 shows, inverting the initial configuration from E/S to S/E did not affect the dynamics and the final result in all the experiments performed with these particles. Initial configuration resulted to have no effect on the segregation propensity of the mixtures analyzed.

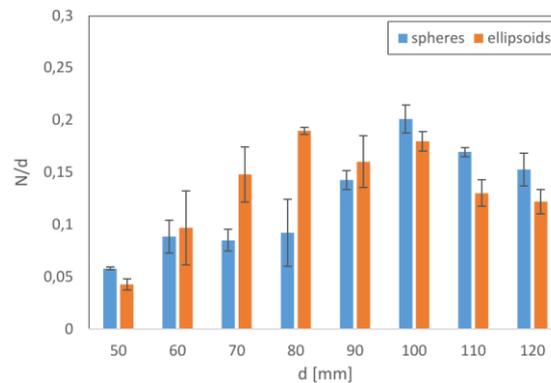

**Figure 5.** Distribution of the radial distance of spheres and ellipsoids with AR=1.5, obtained by averaging the results of three repetitions. Other conditions are t=30 min, $\omega$=4.22 rpm, $\varphi$= 25%.

A more in-depth analysis can be carried out considering the distribution of the radial distance of the spheres and ellipsoids separately. The radial distance from the centre of the drum ($d$) was divided in classes of 10 mm and the number of each type of particles ($N$) in each class was computed. To obtain the distribution, $N$ was normalized on the radial distance as $N/d$ and plotted versus $d$. In the case of completely homogeneous mixtures and assuming that the medium takes the shape of a circular sector, $N/d$ should be constant and equal for the two types of particles in each class. The peak of the distribution of the ellipsoids was found at lower distance than the peak of the distribution of the spheres. Moreover, at lower distance, the number of ellipsoids was higher than the number of spheres, especially for distances between 70 and 90 mm; at higher distance (closer to the drum wall), instead, the number of spheres was higher than the number of ellipsoids, thus confirming the previous analysis based on $d_m$.

Also, binary mixtures made of the two types of ellipsoids E1 and E2 ($AR$=1.5 and $AR$=3.5 respectively) were taken into account at the same filling levels of the previous experimental tests and at the lower rotational speed. Stable segregation was again observed with intermediate results with respect to the two previous cases (S+E1 and S+E2). So as a general comment on these experiments, stable segregation was always observed when mixing particles having same volume but different $AR$. The intensity of segregation increased with increasing the difference in $AR$ between the particles. In particular particles with larger $AR$ tended to accumulate at the center of the granular bed, while particle with lower $AR$ accumulated at the periphery. Segregation occurred in the surface active layer where the more elongated particles, because of their reduced propensity to flow (particle rolling was prevented by their elongated shape), were systematically overtaken by the less elongated (and faster) particles. In this way a spontaneous separation of particles occurred in the active layer, with those having a lower $AR$ arriving first at the drum wall, while those having an higher $AR$ remaining trapped in the core.

## 3.1 Effect of filling level

The comparison of the behavior of the mixtures at the two filling levels can be carried out by normalizing the mean radial distance $d_m$ of the two types of particle on the mean distance of all the particles $d_c$, calculated as $(d_{m,\,spheres}+d_{m,\,ellips.})/2$. The trends of $d_m/d_c$ are shown in Figure 6

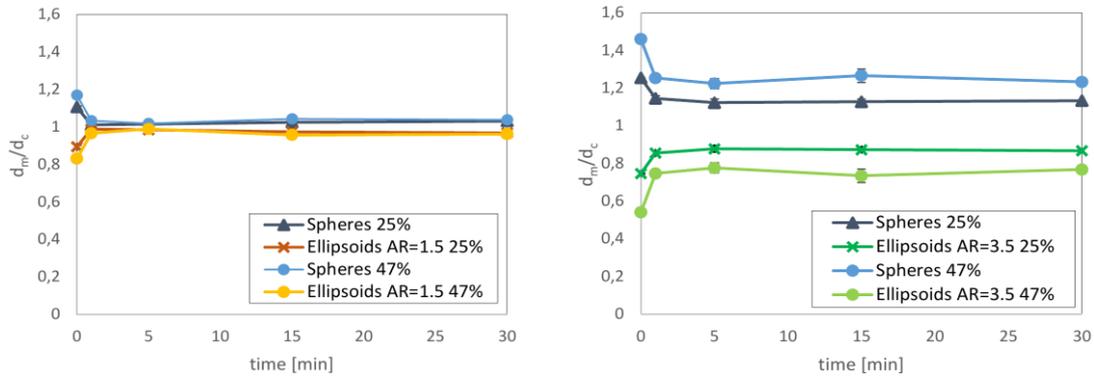

**Figure 6.** Trend of $d_m/d_c$ vs. time for the mixture of spheres and ellipsoids with $AR$=1.5 (left) and $AR$=3.5 (right) at two different filling levels.

In the case of mixture of spheres and ellipsoids with $AR$=1.5, the curves at different filling levels overlap, showing that the behaviour of the mixture was not affected by the filling level of the drum. In the case of the other two mixtures (S+E2 and E1+E2, not shown), the curves did not overlap , especially for the case S+E2 shown in Figure 6 (right). An explanation can be found in terms of bed geometry. When increasing the filling level (below 50%) the length of the surface active layer increased as well. This gave more chances for particles to separate, in particular when the difference in $AR$ was large enough (S+E2 and E1+E2). However for the system S+E1 where the difference in $AR$ was the smallest, the difference in active layer length in the two cases was not sufficient to create observable differences in the segregation pattern.

## 3.2 Effect of rotational speed

In addition to experiments carried out at 4.2 rpm, tests were repeated at 11.5 rpm with spheres and ellipsoids (25% filling level). Figure 7 shows that rotational speed did not affect the segregation patterns, at least in the range of the analyzed speeds.

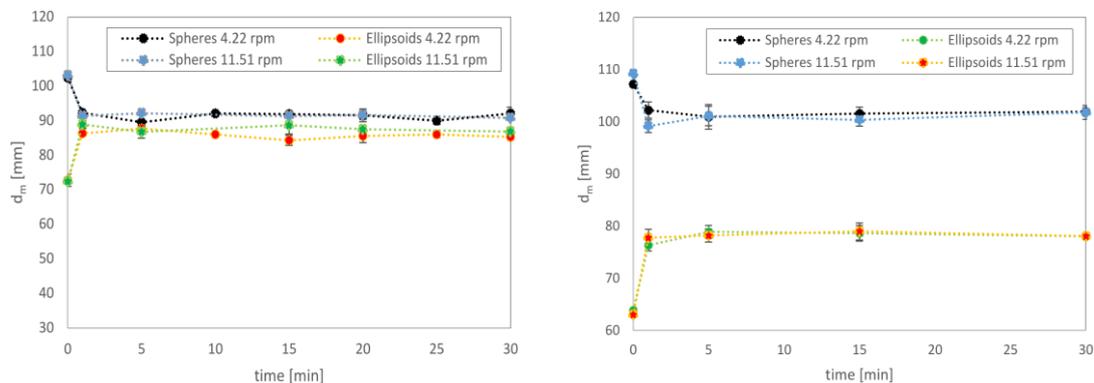

**Figure 7.** Trend of the mean radial distance vs. time for the mixture of spheres and ellipsoids with $AR$=1.5 (left) and $AR$=3.5 (right) at two different rotational speeds.

## Conclusions

From the experimental tests described in the previous paragraphs, it was found that binary mixtures of particles with different shape but with the same volume do not mix uniformly and segregation was always observed. In particular, the particles with lower aspect ratio tended to stay closer to the drum walls, while the more elongated particles were found preferentially at the centre of the granular bed. The most significant case was represented by the mixture of spheres and ellipsoids with $AR$=3.5; for this mixture, in fact, the segregation was more evident and the non-spherical particles clearly accumulated in the centre of the granular bed. The same final state was obtained regardless of whether the longer particles were initially located to the top or the the bottom of the mixture. A proportionality between $AR$ difference and segregation intensity was observed. Some variables such as rotational speed and initial bed configuration were found to not affect the segregation dynamics and intensity. Only filling level (in addition to $AR$) was found to affect segregation intensity. A mechanistic explanation was given in terms of differences in particle flowability occurring in the surface active layer (confirmed also by an analysis of the angles of repose in the drum). In this specific case therefore percolation and inertial effects (typical of size segregation) were not the main driving mechanisms for the observed segregation (the particle indeed had the same volume and similar mass so that sterical and inertial phenomena were minimized). Segregation mainly occurred because of differences in particle velocity in the active layer due to a reduced mobility of particles with larger $AR$.

## References

1. GDR MiDi (2004), "On dense granular flows", The European Physical Journal E, 14, 4, 341–365.
2. Jarray A., Magnanimo V., Ramaioli M., Luding S. (2017) Scaling of wet granular flows in a rotating drum, EPJ Web of Conferences, 140,03078.
3. Volpato S., Canu, P., Santomaso A.C. (2017), "Simulation of free surface granular flows in tumblers", Advanced Powder Technology 28(3), 1028-1037.
4. Mellmann J. (2001), "The transverse motion of solids in rotating cylinders – forms of motion and transition behaviour", Powder Technology, 118, pp 251-270.
5. Windows-Yule C.R.K., Scheper B.J., van der Horn A.J., Hainsworth N., Saunders J., Parker D.J., Thornton A.R. (2016), "Understanding and exploiting competing segregation mechanisms in horizontally rotated granular media", New Journal of Physics, 18, 2, 023013, 2016.
6. Santomaso, A.C., Olivi, M., Canu, P. (2004), "Mechanisms of mixing of granular materials in drum mixers under rolling regime", Chemical Engineering Science, 59 (16), 3269-3280.
7. Kumar N., Magnanimo V., Ramaioli M., Luding S. (2015), "Tuning the bulk properties of bidisperse granular mixtures by small amount of fines", Powder Technology, 293, 94-112.
8. Ramaioli M., Pournin L., Liebling Th. M., (2005), "Numerical and experimental investigation of alignment and segregation of vibrated granular media composed of rods and spheres", Proceedings of the 5th International Conference on Micromechanics of Granular Media, 2, 1359-1363.
9. Santomaso, A.C., Canu, P. (2007), "Single particles properties vs. bulk flowability", Proceedings of 5th International Conference for Conveying and Handling of Particulate Solids.
10. Schneider, C. A., Rasband, W. S., Eliceiri, K. W. (2012), "NIH Image to ImageJ: 25 years of image analysis", Nature methods 9(7): 671-675.